\documentclass[aps,prb,preprint,showpacs,floatfix,superscriptaddress]{revtex4}
\usepackage{amsmath}
\usepackage{graphicx}
\begin{document}

\title{Dynamics of wave packets in two-dimensional crystals under external magnetic and electric fields:
 Vortices formation}

\author{H.\ N.\ Nazareno}
\affiliation{International Center of Condensed Matter Physics,
             Universidade de Bras\'{\i}lia, P.O. Box 04513, 70910-900
             Bras\'{\i}lia - DF, Brazil}
\author{P.\ E.\ de Brito}
\author{E.\ S.\ Rodrigues}
\affiliation{Universidade Cat\'{o}lica de Bras\'{\i}lia,
             Departamento de F\'{\i}sica, Campus \'{A}guas Claras, 72030-070
             Bras\'{\i}lia - DF, Brazil}

\begin{abstract}
In the present work we deal with the dynamics of wave packets in a
two-dimensional crystal under the action of magnetic and electric
fields. The magnetic field is perpendicular to the plane and the
electric field is on the plane. In the simulations we considered a
symmetric gauge for the  vector potential while the initial wave
packet was assumed to have a gaussian structure with given
velocities. The parameters that control the kind of time evolution
of the packets are: the width of the gaussian, its velocity, and,
the intensity and direction of the electric field as well as the
magnitude of the magnetic field. In order to characterize the kind
of propagation we evaluated the mean-square displacement (MSD), the
participation function and which is more important we were able to
follow the wave at different times, which allowed us to see the time
evolution of the centroid of the wave packets. A novel effect was
observed, namely, the dynamics is such that the wave function
\emph{splits} into two or more components and \emph{reconstructs}
successively as time goes, vortices are forming. To our
understanding this is the first time such an effect is reported. As
for the inclusion of the electric field, we observe a complex
behavior of the wave packet as well as note that the vortices
propagate in direction perpendicular to the applied electric field.
A similar behavior presented by the classical treatment, In our case
we give a quantum mechanics explanation for that.

\end{abstract}


\maketitle

\section{Introduction}

In the present work we deal with the problem of  the behavior of
wave packets in a two dimensional square lattice under the action of
magnetic and electric fields. The magnetic field is perpendicular to
the lattice, while the electric field is in the plane.

We can mention the (pioneering) works done on the subject of wave
propagation in low dimensional systems that have attracted the
interest, since the early days of quantum
mechanics.~\cite{bl,re,lo,gh,za}

The subject of carriers in a two dimensional structure under the
action of external magnetic and electric fields, has aroused intense
interest since it has become recently experimentally accessible.
\cite{al,ke,kr,ku}

We have found very interesting properties of the time evolution of
initial wave packets that were assumed of a gaussian structure with
a given velocity. We analyzed gaussians with different dispersions
which in turn determine the type of propagation a wave packet will
present. Another parameter which has a direct influence on the wave
packet behavior is the assumed initial velocity. Obviously, the
magnitude as well as the direction of the electric field has also a
direct influence on propagation. Clearly, one notices the wavy
nature of the solution of the time dependent Schr\"odinger  equation
but, a novel effect is observed, namely, the successive splitting
and reconstruction of the wave function in two or more components as
time goes. We observe the formation of vortices due to the joint
effect of the crystal potential and the external fields. To our
understanding this novel effect has not been reported until now. In
order to comprehend the characteristic of propagation we resort to
the study of the trajectories in reciprocal space since they are
connected with the ones in direct (coordinate) space by a rotation
of $\pi/2$.

A pioneering work dealing with the motion of an electron in a 2D
lattice potential superimposed to a magnetic field is due to Peierls
\cite{re}, that considered an effective single band Hamiltonian
arising from a tight-binding dispersion relation. As a result of
this model the single Bloch band is split into magnetic subbands
according to the number of flux quanta that pierces the unit cell of
the 2D lattice. At the same time, the parameter $\alpha=
\Phi/\Phi_{0}$ being the ratio between the magnetic flux through the
unit cell to the quantum of flux, controls the kind of propagation
of wave packets in a lattice under a uniform magnetic field. For a
rational value we recover translational symmetry, with an enlarged
unit cell which in turn makes possible for a packet to propagate in
the sample. On the contrary for an irrational value of $\alpha$ we
face the problem of incommensurability of the potential which in
turn produces a localization of a wave packet in a definite region
of the lattice~\cite{hp}. Recently, the electronic spectrum of a
two-dimensional quantum dots array under magnetic and electric
fields was presented~\cite{em}.

\section{The Model}
The action of a magnetic field is analyzed along the Peierls
model~\cite{re}, which consists in taking a dispersion relation for
a square lattice
\begin{equation}
E(\textbf{k}) = 2W \left(\,\cos k_{x}a + \cos k_{y}a \, \right)
\label{ek}
\end{equation}
and replacing in it the quasimomentum $\textbf{k}$ by
\begin{equation}
\hbar\textbf{k} \Rightarrow -i\hbar\nabla - e\textbf{A}/c \label{hk}
\end{equation}
to obtain a model Hamiltonian. In the present work we used the
symmetric gauge for the vector potential:
\begin{equation}
{\bf A} =\frac{B}{2} \left(-y\,\hat{\textbf{\i}} +
x\,\hat{\textbf{\j}}+ 0\,\hat{\textbf{k}}\right) \label{vec}
\end {equation}

The classical paper of Hofstadter~\cite{ho} showed that such an
approach leads to a spectrum as function of the magnetic field that
presents a fractal structure, the so called Hofstadter butterfly. On
the other hand, Hall measurements on GaAs/AlGaAs superlattice
provided evidence of the existence of the structure of Hofstadter's
butterfly\cite{ca}. Since this model is of a single band, is limited
to analyze systems of large gaps and/or magnetic intensities such
that no interband transitions occur. Since we want to study the kind
of propagation of a particle in such a system, we expand the wave
function in the Wannier representation:
\begin{equation}
\left|\,\Psi(t)\,\right> = \sum_{mn} \,\,g_{mn}\,(t)\,\,
\left|\,mn\,\right>
\end{equation}
where the ket $\left|\,mn\,\right>$ is the ket associated with the
corresponding site~\cite{lo,gh}. Next, we assume a discrete set of
coordinates such that $x = ma$, and  $y = na$. The time dependent
Schr\"{o}dinger equation in the Wannier representation becomes the
set of equations:
\begin{eqnarray}
i\hbar\,\,\frac{dg_{mn}}{dt}&=& W \left(\,g_{m+1,n}\,\,e^{i\pi\alpha
n} +g_{m-1,n}\,\,e^{-i\pi\alpha n}+ g_{m,n+1}\,\,e^{-i\pi\alpha
m}+g_{m,n-1}\,\,e^{i\pi\alpha m}\,\right)\nonumber\\
& & +\,\, g_{mn}\left(\,\epsilon _{mn}\,+\, e\,E_{x}\,a m \,+\,
e\,E_{y}\,a n \right) \label{Wannier}
\end{eqnarray}
where $\alpha= \Phi/\Phi_{0}$ is the ratio between the flux through
the unit cell in the $(x,y)$ plane to the quantum of flux~\cite{hp}
$\Phi_{0}\,=\,hc/e$,  $\epsilon _{mn}\,$ are the on-site energies
and $W$ is the hopping term.

We have used the Runge Kutta method of fourth order to integrate the
equations of motion. In order to solve the time dependent
Schr\"{o}dinger equation we chose as initial condition a gaussian
wave packet with a certain width and a given velocity:
\begin{equation}
\left\langle \,\,x,y \, \right| \Psi (t=0)\,\,\rangle \,\,=\,\, \exp
i\left( \mathbf{k} \cdot \mathbf{r}\right) \,\,\frac{1}{\sigma
\sqrt{\pi }}\,\,\exp \left[\,\, -
\frac{(x-x_{0})^{2}+(y-y_{0})^{2}}{2\sigma^{2}}\,\,\right]
\end{equation}

We decided for this since it is more realistic to assume that the
injected electron is not extremely localized. One difficulty faced
during the calculations was to decide the right size of the lattice
in order to avoid boundary effects. To be specific, we have taken: a
lattice of $400 \times 400$ sites, $\alpha = 0.0242$ which
corresponds to a magnetic field of intensity $B= 1T$, the
dimensionless units of time $\tau = Wt/\hbar$, corresponds to $0.026
ps$, the dimensionless units of electric field are $W/ea$, such a
unit is equivalent to $12.5 \, kV/cm$, and all lengths are in units
of the lattice parameter $a$. After solving the set of equations we
constructed the following:

i) the mean square displacement $(MSD)$
\begin{equation}
<{\textbf{r}^{2}>(t) \,\,=\,\, \sum_{mn}\,\,|\,g_{mn}(t)\,|^{2}
\,\,\left(\,n^{2}a^{2}\,+\, m^{2}a^{2}\,\right)} \label{msd}
\end{equation}

ii) we follow the centroid of the wave packet by evaluating the
following quantity, which give us the amount of the displacement
form the initial position of the particle:
\begin{equation}
<\Delta x>(t) \,=\, \,\sum_{n}\,(n-n_{0})\,\,
\sum_{m}|\,\,g_{mn}(t)\,\,|^{2}\,\,
\end{equation}

\begin{equation}
<\Delta y>(t) \,=\, \,\sum_{m}\,(m-m_{0})\,\,
\sum_{n}|\,\,g_{mn}(t)\,\,|^{2}\,\,
\end{equation}

iii) the participation function~\cite{we}
\begin{equation}
P(t) \,=\, \left[\,\sum_{mn}\,\, |\,g_{mn}(t)\,|^{4}\,\,\right]^{-1}
\label{pt}
\end{equation}
An interesting feature of this function is that it indicates the
sites that participate in the wave packet. At the same time it
presents an abrupt decline once the packet reaches the boundary of
the lattice, in this way we can note the presence of size effects.
We followed the Anderson~\cite{an} criterium for analyzing
diffusion, namely, we can conclude that diffusion has occurred if at
$ t \rightarrow \infty$ the Wannier amplitudes at the starting sites
go to zero. If, these amplitudes remain finite decreasing rapidly
with distance, we say we have a localized state. We also plot the
wave packet as it evolves in time which tells us the kind of
propagation for the different cases in study.  More than that, by
looking at the displacements of the maxima of the packet we can
infer the kind of \textit{trajectory} a particle would describe.
Besides that, we follow the time evolution of the centroid of the
wave packet which gives complementary information.

\section{The splitting of the wave packet: Vortices formation}

We would like at this point to describe a surprising behavior
observed during the time evolution of wave packets. First, we note
that their gaussian structure means that in reciprocal space we have
a certain dispersion in $\textbf{k}$, which implies that several
wave vectors will participate in the evolution of the wave function.
This plays an important role as long as we are considering the wave
vector associated with the velocity of the initial wave packet,
lying along and around the lines of zero energy in the Brillouin
zone of the square lattice. In Fig. 1 we show the lines of constant
energy where the arrows signal the orbit in reciprocal space
described by the wave vector.

As it is well described in the books of Solid State\cite{ki}, the
quasimomentum satisfies an equation of motion analogous to the
classical one; where one takes the group velocity $ \textbf{v}=
\frac{1}{\hbar}\frac{\partial \varepsilon}{\partial \textbf{k}}$ in
the expression of the Lorentz force, which in turn determines that
the wave vector moves along the lines of constant energy. First,
consider $\textbf{k}$ \textit{inside} the region of energy zero, but
close to its boundaries, it will describe a clockwise orbit in
reciprocal space. As for $\textbf{k}$ \textit{outside} but close to
the line, we get another trajectory described in counterclockwise
sense. This will result in the appearance of vortices rotating in
opposite directions when describing the evolution of the wave packet
in direct space, where the "orbits" are obtained after rotation by
$\pi/2$.

What is very interesting is to consider $\textbf{k}$ on one of the
lines of zero energy. In such a case due to the dispersion in
$\textbf{k}$ because of the gaussian structure of the initial
packet, we discussed above, we have to take into account
$\textbf{k}$ values \emph{inside} and \emph{outside} the line,
symmetrically distributed. Consequently, part of the wave (half of
it) will describe a trajectory in the clockwise sense and the other
half of the wave in the opposite sense, the resulting movements is
similar to a swimmer doing breaststroke. As time goes, the wave
packet starts moving in the direction of the velocity and then
splits into two components that will join in a single component and
so on. Consequently, two vortices with opposite angular velocities
are formed. The wave remains stationary in a certain region of the
lattice. See Fig. 2 where this remarkable effect is shown. Assuming
now $\textbf{k}$ with components ($\pi/a, 0$), as shown in Fig. 1,
four squares in reciprocal space participate in the movement of the
packet reinforcing each other such that at a certain time the wave
appears like is shown in Fig. 3. To get the evolution of the wave
packet one has to follow the arrows in each of the four regions.
Again, we observe the splitting and reconstruction of the wave, this
time in several components. In this case, several vortices are
present.

One encounters a similar effect when considering a 1-D crystal under
the action of a dc electric field. When considering as an initial
state a well localized state at a site in the system, the evolution
of the packet is such that it is split into two symmetric parts,
which oscillate with the Bloch frequency in opposite
directions\cite{hn}.

Let us assume $\textbf{k}$ inside the region limited by the lines of
zero energy but close to one of them. The wave function will split
into two components in such a way that one part of the wave, the
major part, will rotate clockwise while the rest will do
counterclockwise. As for $\textbf{k}$ outside the region but close
to one of the lines, the reciprocal is true, the major part of the
wave will perform a rotation counterclockwise. To illustrate this
effect, consider now the wave vector of the initial wave packet,
inside the region of the lines for $\epsilon =0$ but close to one of
it, we took for example $\textbf{k}a = (1.4, 1.4)$. As said above,
the wave is split in two components, where the bigger part rotates
clockwise and the smaller part rotates in the opposite sense. In
this case this case it results in the appearance of two asymmetric
vortices.  See Fig. 4.

\section{The effect of a dc electric field}

We consider now the inclusion of a dc electric field in the equation
of motion for the Wannier amplitudes. As a general trend we observe
that the packet will propagate in a more complex way, but
\emph{always} in direction perpendicular to the electric field, as
it was shown previously\cite{hp}. This behavior is also present in
the classical treatment of the problem\cite{la}. From the view point
of Quantum Mechanics, we understand this behavior since the electric
field breaks the degeneracy of the on-site energies along the
direction of the applied field, inhibiting hopping between these
sites. First we take $\textbf{k}$ on one of the lines of zero
energy, for example ($\pi/2a, \pi/2a$) and the electric field along
the diagonal, $\textbf{E}=(0.1, 0.1)$, while the initial packet has
$\sigma = 1$. The wave is split while propagating but, due to the
presence of the electric field, one part of the wave proceeds with a
greater velocity, with a centroid trajectory similar to the one a
\emph{snake} performs when moving around. The other part that moves
in opposite direction, remains close to the starting point. See Fig.
5 where we also show the centroid trajectory, and the MSD and
participation as functions of time.

For the same configuration but for $\sigma = 2$, we observe that the
wave is \emph{not} split and follows a trajectory quite similar to
the classical one, i.e., the more so, the more extended is the
initial wave packet, i.e., the greater is sigma. This comes about
since a greater dispersion in direct space is related to a smaller
one in reciprocal space. See Fig. 6. For the case $\sigma = 3$ we
confirm this behavior of the propagation of the wave. It is
interesting to mention that the trajectories for $\sigma=2$ and $3$
are exactly the same. By increasing the intensity of the electric
field we obtain a displacement of the centroid along a trajectory
similar to the former one, the only difference is that in the case
of stronger field the amplitude of the displacement as well as the
period of the oscillations are reduced. This field effect is shown
at the bottom left of Fig. 6.

A very peculiar effect is obtained by considering the electric field
with components $(-0.1, 0.1)$  and taken $\textbf{k}= (\pi/2a,
\pi/2a)$. In this case we note that the "swimmer" by doing
breaststroke can displace itself along the perpendicular direction
to the applied electric field. The vortices displacement is such
that the centroid moves along a \emph{strait} line, as shown in
Fig.7. This should be compared with the case without electric field
in which the packet remains stationary as shown in Fig. 2.

\section{Conclusions}
We show in this work a novel effect of wave packet propagation in a
two dimensional crystalline system under the action of combined
magnetic and electric fields. Using the Runge Kutta method of forth
order, we integrate the equations of motion in the Wannier
representation, assuming as initial condition a gaussian wave packet
with given velocity. We found the very interesting behavior of the
wave function, namely, that by taking the initial velocity with the
associated wave vector close to the lines of zero energy, the wave
is split and reconstructed as time goes, with the appearance of a
series of vortices. This effect comes about since we used a gaussian
as an initial condition. This in turn, implies that one has to take
into account a dispersion in the reciprocal space, so there are
contributions of the lines of  constant energy on both sides of the
zero energy line, as explained above. Without the presence of the
electric field, the wave remains in a definite region of the
lattice. The inclusion of the electric field, produces a
displacement of the vortices along the perpendicular direction of
the applied field, showing a more complex behavior since now,
besides the displacement, the wave is split and reconstructed as
time goes. As for the centroid trajectory, for a configuration such
that the applied electron field is parallel to the initial velocity,
it describes a trajectory similar to the way a snake performs when
moving around. In the classical treatment, the trajectory is a
trochoid. For other configurations, i.e., for the wave packet with
initial velocity perpendicular to the field, the centroid trajectory
is a strait line. One last comment deserves to be made, the effects
we described are the results of taking into account that the
particle, besides being under the action of the fields, is subjected
to a 2D crystal potential as well.

\newpage

\begin{figure}[htbp]
\centerline{\includegraphics[width=16cm,angle=0]{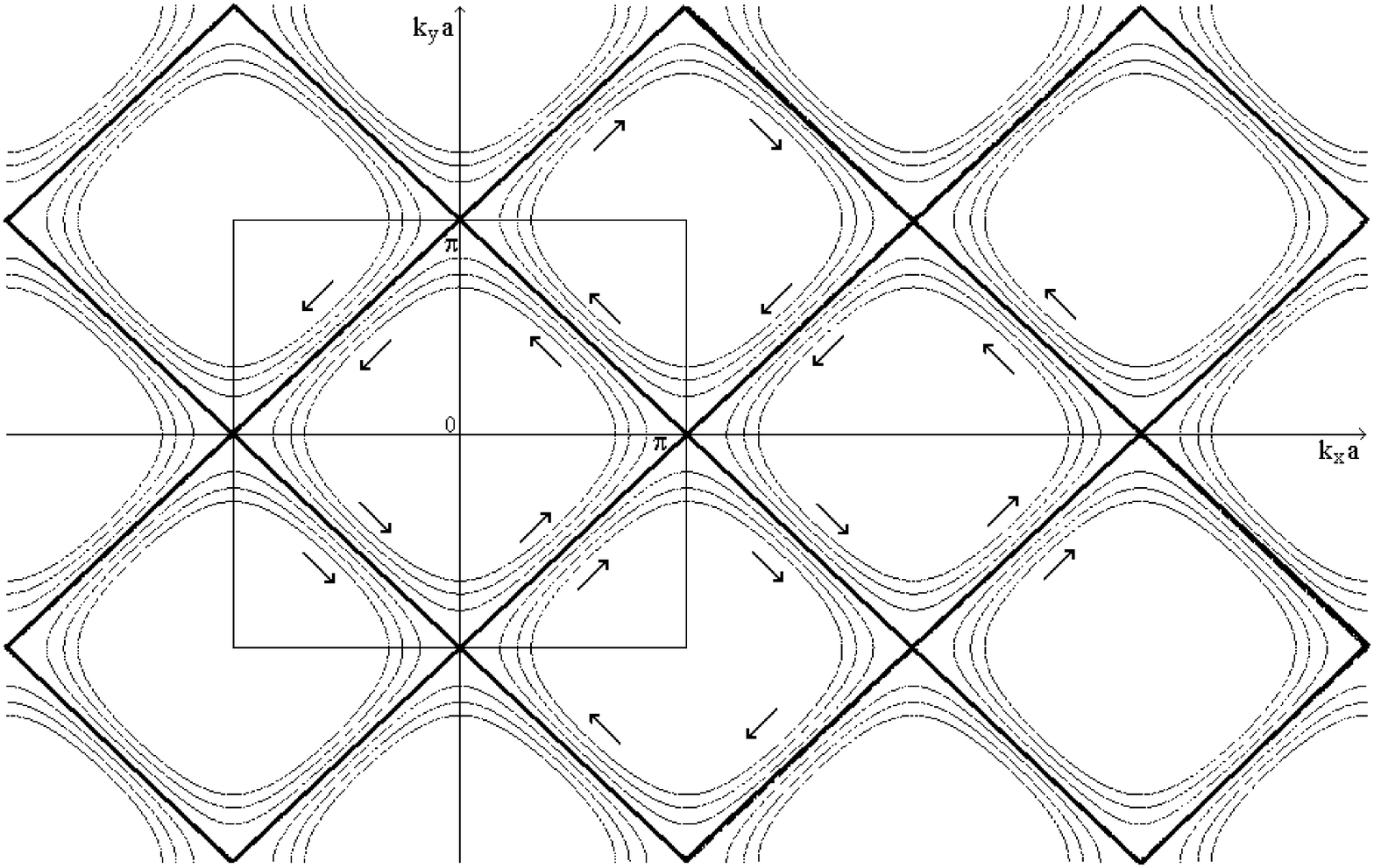}}
\caption{ We show the lines of  constant energy for the square
lattice. The solid lines are the ones corresponding to zero energy.
The arrows indicate the orbits in reciprocal space.} \label{Fig1}
\end{figure}

\begin{figure}[htbp]
\centerline{\includegraphics[width=18cm,angle=0]{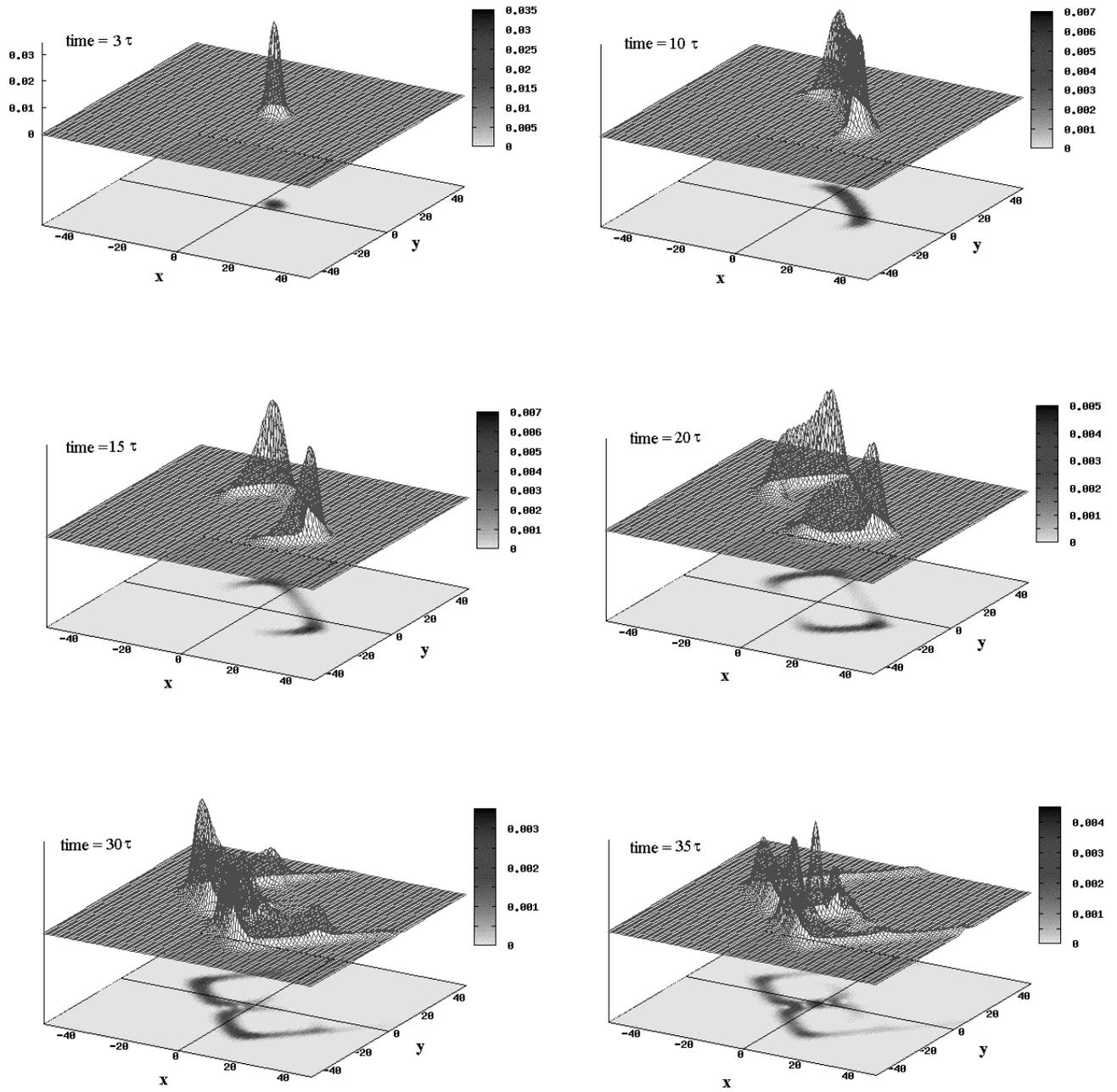}}
\caption{The time evolution of the wave packet for the following
parameters: $\alpha = 0.0242$, dispersion $\sigma = 2$, and $
\textbf{k} a = (\pi/2, \pi/2)$. } \label{Fig2}
\end{figure}

\begin{figure}[htbp]
\centerline{\includegraphics[width=18cm,angle=0]{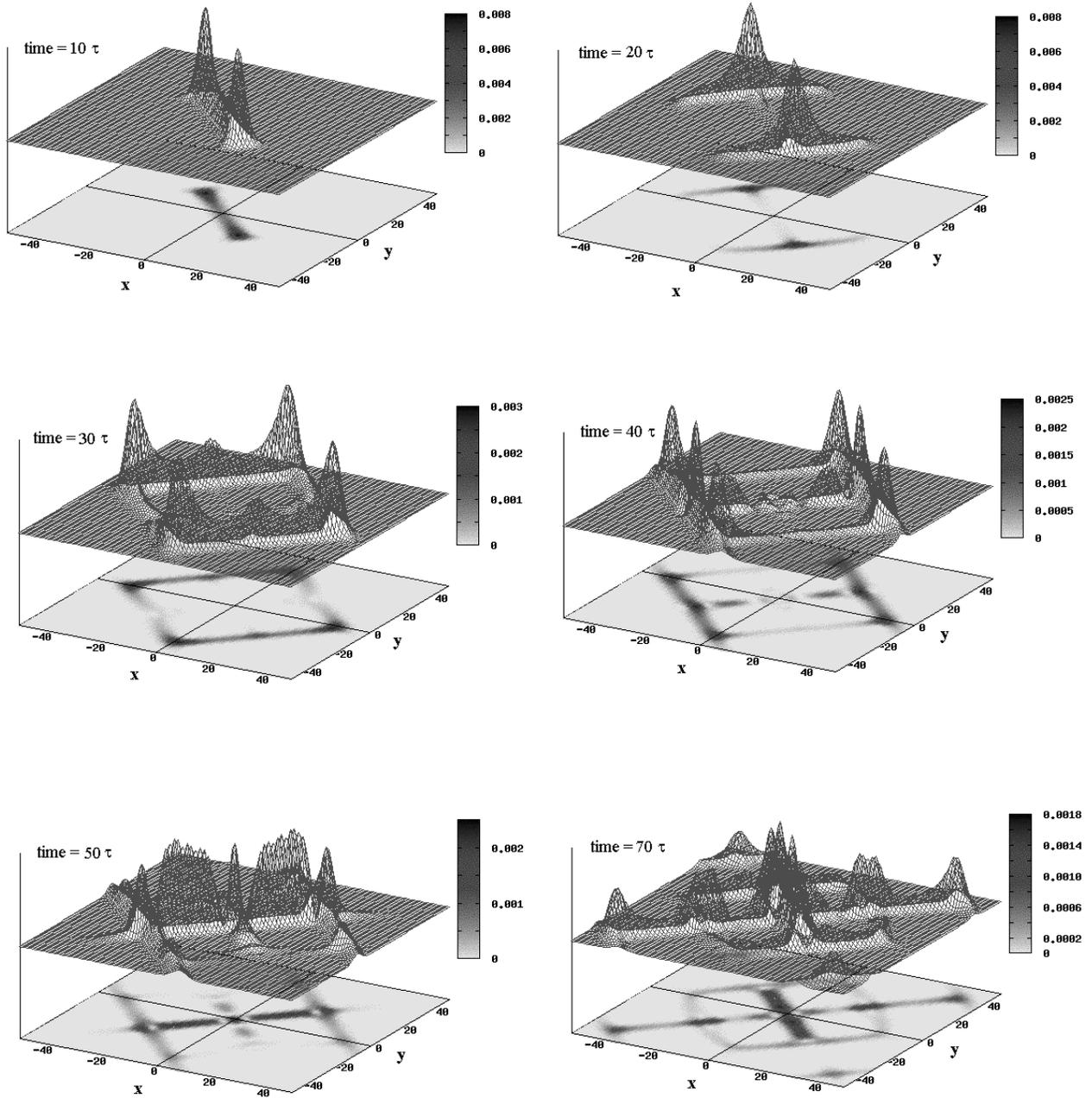}}
\caption{The same as Fig. 2 but for $ \textbf{k} a = (\pi, 0)$. }
\label{Fig3}
\end{figure}

\begin{figure}[htbp]
\centerline{\includegraphics[width=18cm,angle=0]{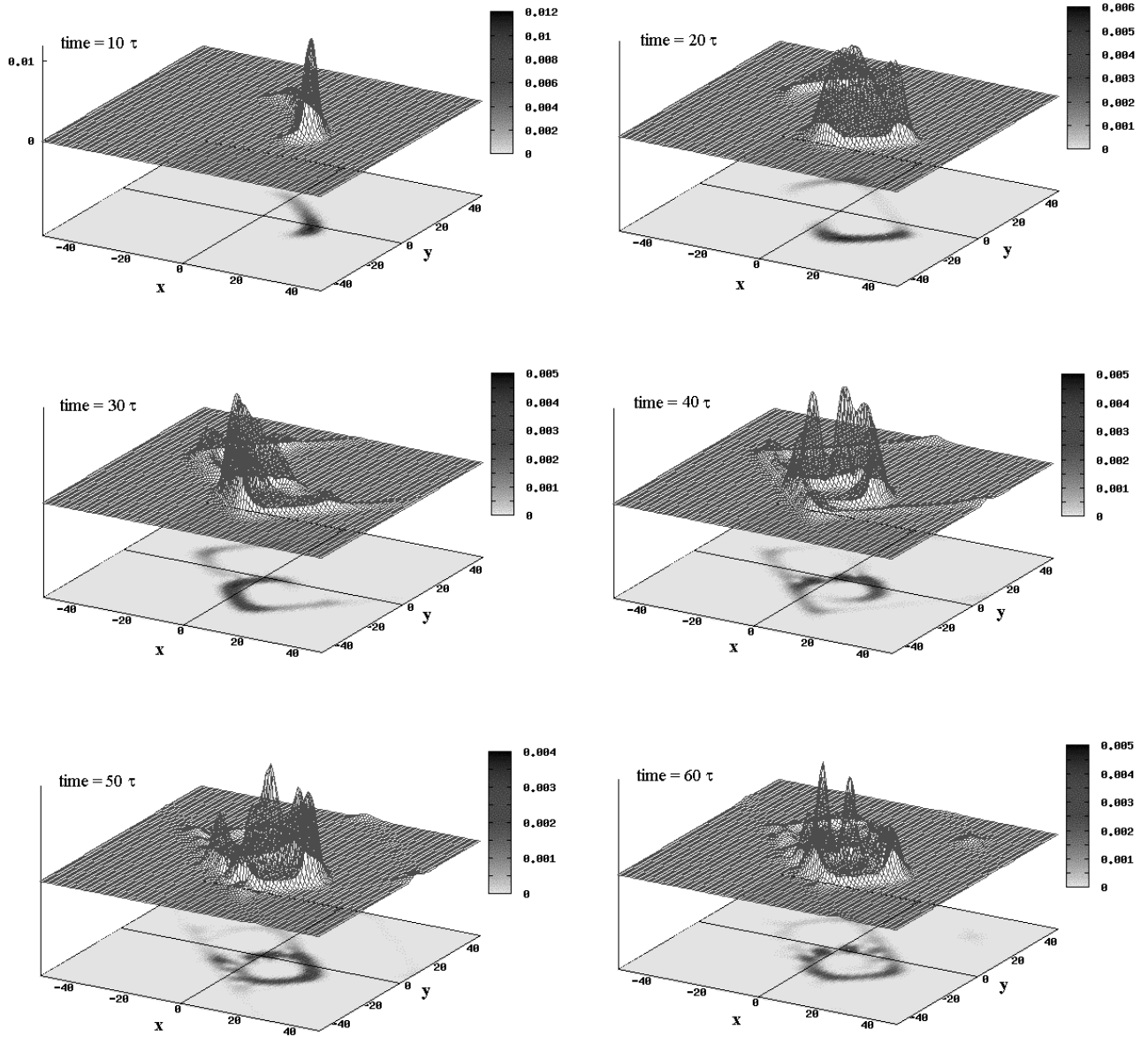}}
\caption{The same as Fig. 3 but for $ \textbf{k} a = (1.4, 1.4)$. }
\label{Fig4}
\end{figure}

\begin{figure}[htbp]
\centerline{\includegraphics[width=18cm,angle=0]{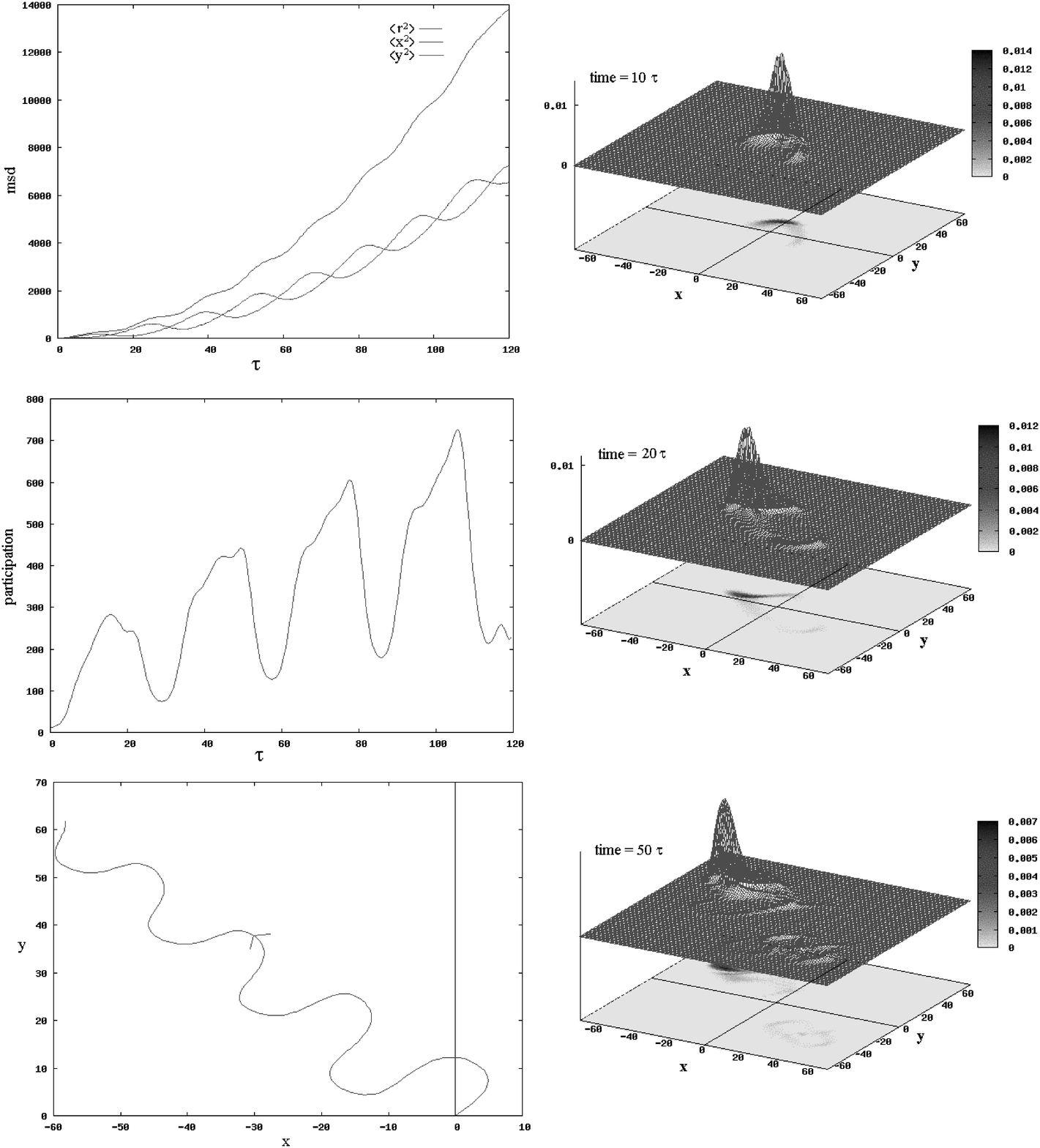}}
\caption{The time evolution of the wave packet for the following
parameters: $\textbf{E}= (0.1, 0.1)$, $\sigma = 1$ and $ \textbf{k}
a = (\pi/2, \pi/2)$. On the left, MSD and participation as functions
of time, and the centroid trajectory which resembles a trochoid. }
\label{Fig5}
\end{figure}

\begin{figure}[htbp]
\centerline{\includegraphics[width=18cm,angle=0]{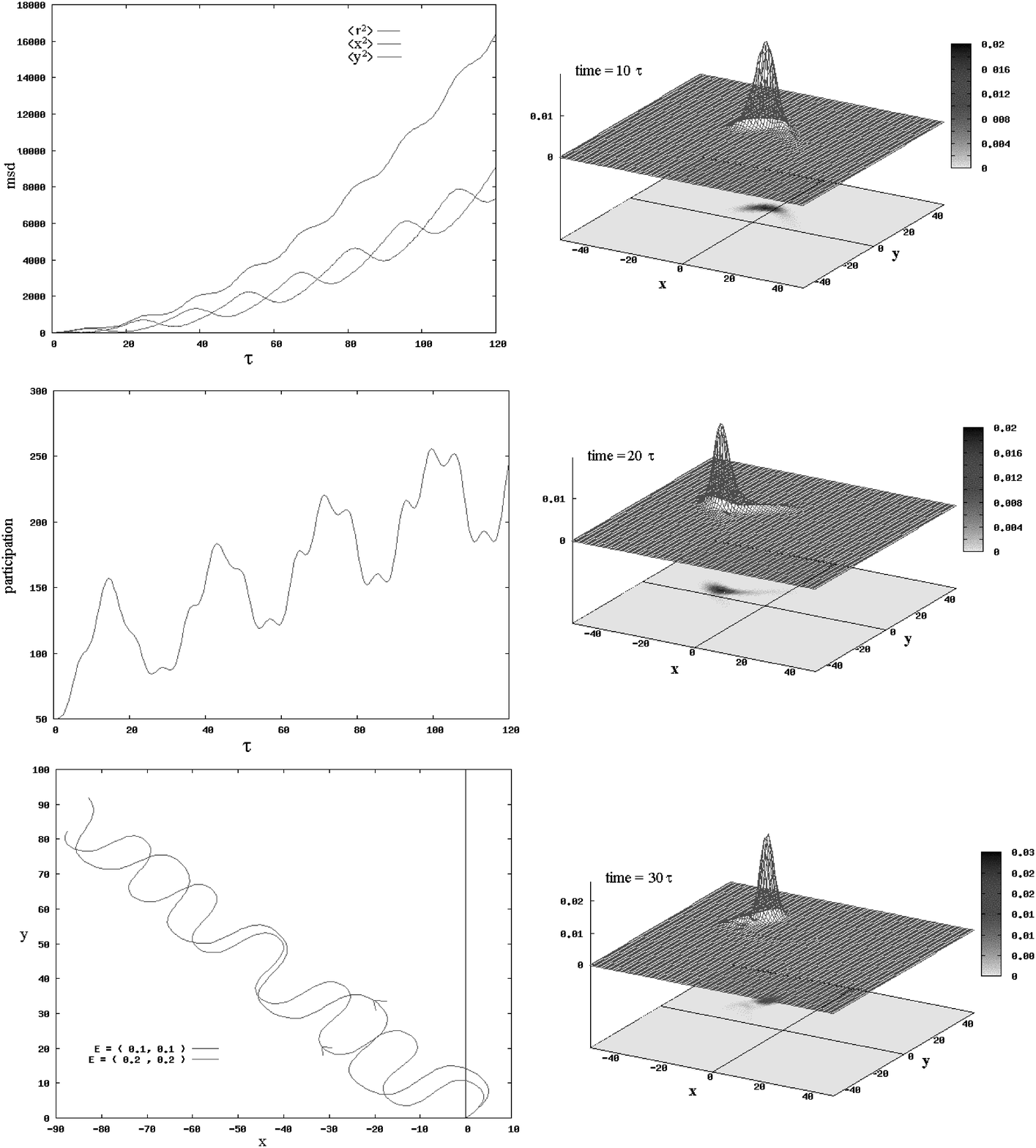}}
\caption{The same as Fig. 5 but for $\sigma = 2$. The bottom left
shows the trajectories of the centroids for two field intensities:
$\textbf{E}= (0.1, 0.1)$, and $\textbf{E}= (0.2, 0.2)$. Note the
different periods when varying the field intensity. See the text. }
\label{Fig6}
\end{figure}

\begin{figure}[htbp]
\centerline{\includegraphics[width=18cm,angle=0]{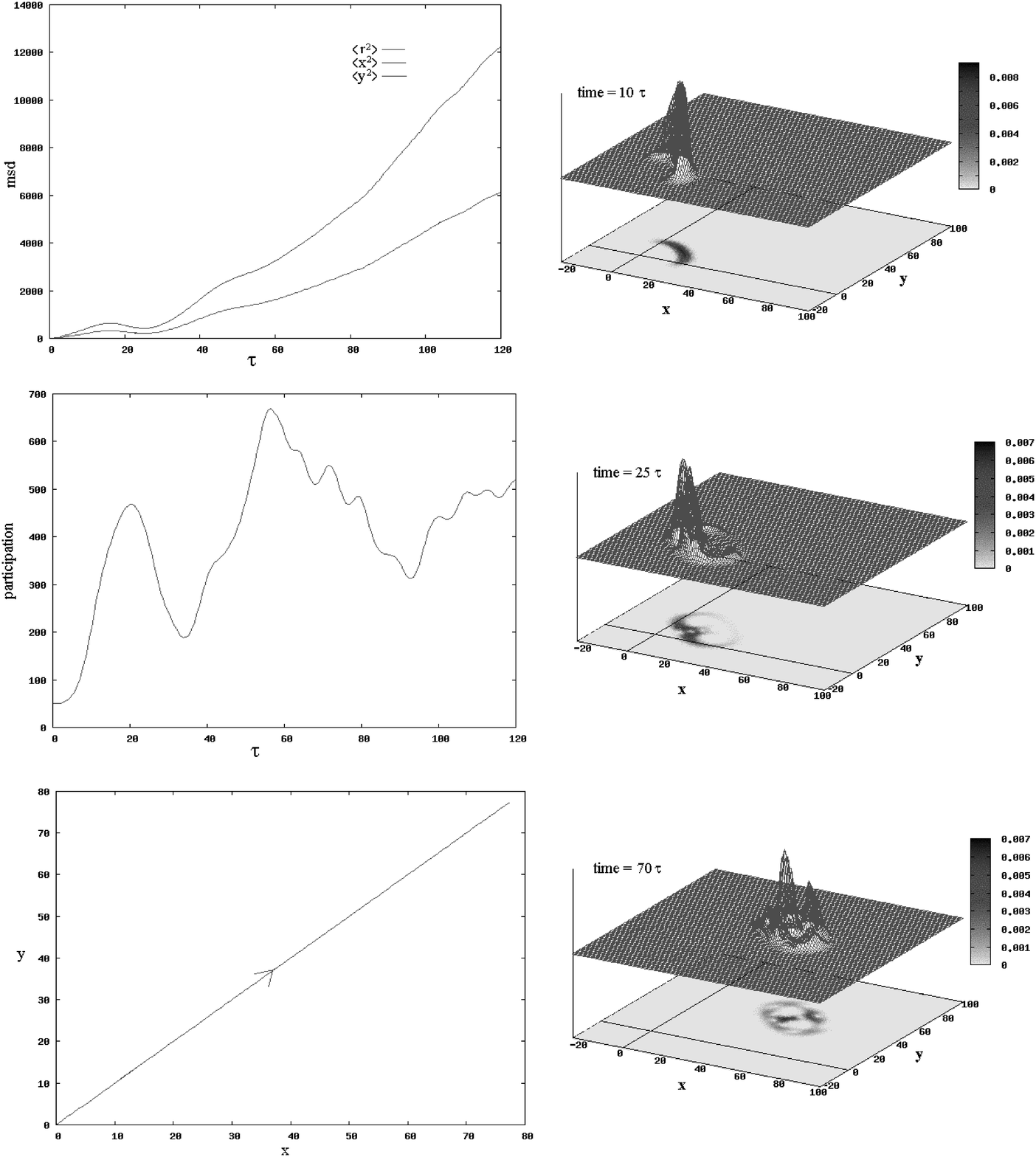}}
\caption{The same as Fig. 6 but for $\textbf{E}= (-0.1, 0.1)$. Note
the strait line described by the centroid. } \label{Fig7}
\end{figure}

\end{document}